\documentclass{tlp}

\newtheorem{theorem}{Theorem} 
 
\newtheorem{example}[theorem]{Example}

\usepackage{rotating}
\usepackage{subfigure}
\usepackage{amssymb}
\usepackage{amsmath}
\usepackage{url}
\usepackage[all]{xy}
\usepackage{color}
\usepackage{sudoku}
\usepackage{rotating}
\usepackage{listings}
\usepackage{graphics}
\usepackage{graphicx}
\usepackage{wrapfig}
\usepackage{enumitem}

\newcommand{\true}{\texttt{true}}

\newcommand{\allDifferent}{\texttt{allDiff}}

\newcommand{\sset}[2]{\left\{~#1  \left|
      \begin{array}{l}#2\end{array}
    \right.     \right\}}
\newcommand\tuple[1]{\langle #1 \rangle}

\newenvironment{SProg}
     {\begin{small}\begin{tt}\begin{tabular}[c]{l}}%
     {\end{tabular}\end{tt}\end{small}}

\newcommand{\false}{\mathit{false}}
\renewcommand{\true}{\mathit{true}}

\newcommand{\ignore}[1]{}

\newcommand{\bee}{\textsf{BEE}}

\begin{document}
\title[Compiling Finite Domain Constraints]{ Compiling Finite Domain
  \\Constraints to SAT with \bee
}

\author[A.~Metodi and M.~Codish]{AMIT METODI\qquad MICHAEL CODISH\\
Department of Computer Science, Ben-Gurion University, Israel
}

 \submitted {25 March 2012}
 \revised   {11 June  2012}
 \accepted  {18  June  2012}

\maketitle
\label{firstpage}

\begin{abstract}
  We present \bee, a compiler which enables to encode finite domain
  constraint problems to CNF. Using \bee\ both eases the encoding
  process for the user and also performs transformations to simplify
  constraints and optimize their encoding to CNF. These optimizations
  are based primarily on equi-propagation and on partial evaluation,
  and also on the idea that a given constraint may have various
  possible CNF encodings. Often, the better encoding choice is made
  after constraint simplification. \bee\ is written in Prolog and
  integrates directly with a SAT solver through a suitable Prolog
  interface. We demonstrate that constraint simplification is often
  highly beneficial when solving hard finite domain constraint
  problems. A  \bee\ implementation is available with this
  paper. 
\end{abstract}



\begin{keywords}
 SAT encoding,
 FD constraints, Equi-propagation, partial evaluation.
\end{keywords}

\section{Introduction}

In recent years, Boolean SAT solving techniques have improved
dramatically. Today's SAT solvers are considerably faster and able to
manage larger instances than yesterday's. Moreover, encoding and
modeling techniques are better understood and increasingly
innovative. SAT is currently applied to solve a wide variety of hard
and practical combinatorial problems, often outperforming
dedicated algorithms.
The general idea is to encode a (typically, NP) hard problem instance,
$\mu$, to a Boolean formula, $\varphi_\mu$, such that the solutions of
$\mu$ correspond to the satisfying assignments of $\varphi_\mu$. Given
the encoding, a SAT solver is then applied to solve $\mu$.

Tailgating the success of SAT technology are a variety of tools which
can be applied to specify and then compile problem instances to
corresponding SAT instances. 
For example, \citeN{npspec2005} introduce \textsf{NP-SPEC}, a
logic-based specification language which allows to specify
combinatorial problems in a declarative way. At the core of this
system is a compiler which translates specifications to CNF formula.
The general objective of such tools is to facilitate the process of
providing high-level descriptions of how the (constraint) problem at
hand is to be solved.  Typically, a constraint based modeling language
is introduced and used to model instances.  Drawing on the analogy
to 
programming languages, given such a description, a
compiler then provides a low-level executable for the underlying
machine. Namely, in our context, a formula for the underlying SAT or
SMT solver.
One obstacle when seeking to optimize CNF encodings derived from
high-level descriptions, is that CNF encodings are
\textit{``bit-level''} representations and do not maintain
\textit{``word-level''} information. For example, from a CNF encoding
one cannot know that certain bits originate from the same integer
value in the original constraint. This limits the ability to apply
optimizations which rely on such word-level information.

We mention two relevant tools.
Sugar \cite{sugar2009}, is a SAT-based constraint solver. To solve a
finite domain linear constraint satisfaction problem it is first
encoded to a CNF formula by Sugar, and then solved using the MiniSat
solver \cite{minisat2003}. \bee\ is like Sugar, but applies
optimizations. Sugar is the first system which demonstrates the
advantage in adopting the, so-called, unary order-encoding to
represent integers.  We follow suite, and introduce additional novel
encoding techniques that take advantage of, previously unobserved,
properties of the order-encoding.
MiniZinc~\cite{miniZinc2007}, is a constraint modeling language which
is compiled by a variety of solvers to the low-level target language
FlatZinc for which there exist many solvers. It creates a standard for
the source language (which we follow loosely).  \bee\ is like FlatZinc,
but with a focus on a subset of the language relevant for finite
domain constraint problems.

We present a tool, \bee\ {\small (\textbf{\textsf{B}}en-Gurion
  University \textbf{\textsf{E}}qui-propagation
  \textbf{\textsf{E}}ncoder)} which translates models in a constraint
based modeling language, similar to Sugar and FlatZinc, to CNF.
Conceptually, \bee\ maintains two representations for each constraint
in a model so that each constraint is also viewed as a Boolean
function. Partial evaluation, and other word-level techniques, drive
simplification through the constraint part; whereas, equi-propagation
\cite{Metodi2011}, and other bit-level techniques, drive
simplification through the Boolean part.
Finally, an encoding technique is selected for a constraint, depending
on its context, to derive a CNF.

The name, ``\bee'' refers both to the constraint language as well
as to its compiler to CNF. \bee\ is not a constraint solver, but can
be applied in combination with a SAT solver to solve finite domain
constraint problems.
We report on our experience with applications which indicates that
using \bee, like any compiler, has two main advantages. On the one
hand, it facilitates the process of programming (or modeling). On the
other hand, given a program (a model), it simplifies the corresponding
CNF which, in many cases, is faster to solve than with other
approaches.  The tool integrates with SWI Prolog and can be downloaded
from~\cite{bee2012}.

\section{Representing Integers}

A fundamental design choice when encoding finite domain constraints
concerns the representation of integer variables.
\citeN{DBLP:conf/cp/Gavanelli07} surveys several of the possible
choices (the \emph{direct-}, \emph{support-} and \emph{log-}
\emph{encodings}) and introduces the \emph{log-support encoding}.
We focus in this paper on the use of unary representations and
primarily on the, so-called, \emph{order-encoding} (see
e.g.~\cite{baker,BailleuxB03}) which has many nice properties when
applied to small finite domains. We describe the setting where all
integer variables are represented in the order-encoding except for
those involved in a global ``all-different'' constraint which take a
dual representation with channeling between the order-encoding and the
\emph{direct encoding}. This choice derives from the observation by
\citeN{direct4allDiff} that the direct-encoding is superior when
encoding the all-different constraint.

Let bit vector $X=[x_1,\ldots,x_n]$ represent a finite domain integer
variable.  In the \emph{order-encoding}, $X$ constitutes a monotonic
non-increasing Boolean sequence. Bit $x_i$ is interpreted as $X\geq
i$.  For example, the value 3 in the interval $[0,5]$ is represented
in 5 bits as $[1,1,1,0,0]$.  In the \emph{direct-encoding}, $X$
constitutes a characteristic function (exactly one bit takes value 1)
and $x_i$ is interpreted as stating $X= i-1$.  For example, the value
3 in the interval $[0,5]$ is represented in 6 bits as $[0,0,0,1,0,0]$.

An important property of a Boolean representation for finite domain
integers is the ability to represent changes in the set of values a
variable can take. 
It is well-known that the order-encoding facilitates the propagation
of bounds. Consider an integer variable $X=[x_1,\ldots,x_n]$ with
values in the interval $[0,n]$.  To restrict $X$ to take values in the
range $[a,b]$ (for $1\leq a\leq b\leq n$), it is sufficient to assign
$x_{a}=1$ and $x_{b+1}=0$ (if $b<n$). The variables $x_{a'}$ and
$x_{b'}$ for $0\geq a'> a$ and $b<b'\leq n$ are then determined true
and false, respectively, by \emph{unit propagation}.  For example, 
given $X=[x_1,\ldots,x_9]$, assigning $x_3=1$ and $x_6=0$ propagates
to give $X=[1,1,1,x_4,x_5,0,0,0,0]$, signifying that
$dom(X)=\{3,4,5\}$.
This property is exploited in Sugar \cite{sugar2009} which also
applies the order-encoding.

We observe, and apply in \bee, an additional property of the
order-encoding: its ability to specify that a variable cannot take a
specific value $0\leq v\leq n$ in its domain by equating two
variables: $x_{v}=x_{v+1}$.
This indicates that the order-encoding is well-suited not only to
propagate lower and upper bounds, but also to represent integer
variables with an arbitrary, finite set, domain.
For example, given $X=[x_1,\ldots,x_9]$, equating $x_2=x_3$ imposes
that $X\neq 2$. Likewise $x_5=x_6$ and $x_7=x_8$ impose that $X\neq 5$
and $X\neq 7$. Applying these equalities to $X$ gives,
$X=[x_1,\underline{x_2,x_2},x_4,\underline{x_5,x_5},\underline{x_7,x_7},x_9]$,
signifying that $dom(X)=\{0,1,3,4,6,8,9\}$.

The order-encoding has many additional nice features that are
exploited in \bee\ to simplify constraints and their encodings to
CNF. To illustrate one, consider a constraint of the form
$\mathtt{A+B=5}$ where \texttt{A} and \texttt{B} are integer values in
the range between 0 and 5 represented in the order-encoding. At the
bit level we have: $\mathtt{A=[a_1,\ldots,a_5]}$ and
$\mathtt{B=[b_1,\ldots,b_5]}$.  The constraint is satisfied precisely
when $\mathtt{B=[\neg a_5,\ldots,\neg a_1]}$. Instead of encoding the
constraint to CNF, we substitute the bits $\mathtt{b_1,\ldots,b_5}$ by
the literals $\mathtt{\neg a_5,\ldots,\neg a_1}$, and remove the
constraint. In Prolog, this is implemented as a unification and does
not generate any clauses in the encoding.

\section{Constraints in \bee}

 
Boolean constants ``$\true$'' and ``$\false$'' are viewed as
(integer) values ``1'' and ``0''.
Constraints are represented as (a list of) Prolog terms. Boolean and
integer variables are represented as Prolog variables, which may be
instantiated when simplifying constraints.
Table~\ref{tab:beeStntax} introduces the syntax for (a simplified
subset of) \bee. In the table, $\mathtt{X}$ and $\mathtt{Xs}$ (possibly
with subscripts) denote a literal (a Boolean variable  or 
its negation) and a vector of literals, 
$\mathtt{I}$ (possibly with subscript)
denotes an integer variable, and $\mathtt{c}$ (possibly with
subscript) denotes an integer constant.

\begin{table}[t]
  \centering
\begin{tabular}{rlll}
\hline\hline
\multicolumn{4}{l}{\bf\small Declaring Variables}\\
\hline
(1) &$\mathtt{new\_bool(X)}$ &\qquad\qquad &declare Boolean \texttt{X}
\\
(2) &$\mathtt{new\_int(I,c_1,c_2)}$ & & 
               declare integer \texttt{I}, $\mathtt{c_1\leq I\leq c_2}$ \\
(3) &    $\mathtt{ordered([X_1,\ldots,X_n])}$ &
          &
          $\mathtt{X_1\geq X_2\geq\cdots\geq X_n}$ (on Booleans)\\
\hline
\multicolumn{4}{l}{\bf\small Boolean (reified) Statements~ 
       \hfill $\mathtt{op\in\{or, and, xor, iff\}}$}\\
\hline
(4) &    $\mathtt{bool\_eq(X_1,X_2)}$ ~or~ $\mathtt{bool\_eq(X_1,-X_2)}$&
          $\mathtt{}$&
          $\mathtt{X_1 = X_2}$ ~or~ $\mathtt{X_1 = \neg X_2}$\\
(5) &    $\mathtt{bool\_array\_op([X_1,\ldots,X_n])}$ &
          $\mathtt{}$&
          $\mathtt{X_1 ~op~ X_2 \cdots op~ X_n}$\\
(6) &    $\mathtt{bool\_array\_op\_reif([X_1,\ldots,X_n],~X)}$ &
          $\mathtt{}$&
          $\mathtt{X_1 ~op~ X_2 \cdots op~ X_n\Leftrightarrow X}$\\
(7) &    $\mathtt{bool\_op\_reif(X_1,X_2,~X)}$ &
          $\mathtt{}$&
          $\mathtt{X_1 ~op~ X_2\Leftrightarrow X}$\\
(8) &     $\mathtt{bool\_array\_lex(Xs_1,Xs_2)}$ &$\mathtt{}$&
             $\mathtt{Xs_1}$ precedes $\mathtt{Xs_2}$ in the lex order\\
\hline

\multicolumn{4}{l}{\bf\small Integer relations (reified)
                    \hfill $\mathtt{rel\in\{leq, geq, eq, lt, gt, neq\}}$}\\
\multicolumn{4}{l}{\bf\small and arithmetic \hfill
            ~$\mathtt{op\in\{plus, times, div, mod, max, min\}}$, 
             $\mathtt{op'\in\{plus, max, min\}}$  }\\
    \hline
(9) &     $\mathtt{int\_rel(I_1,I_2)}$ &
          $\mathtt{}$&
          $\mathtt{I_1 ~rel~ I_2}$\\
(10) &    $\mathtt{int\_rel\_reif(I_1,I_2,~X)}$ &
          $\mathtt{}$&
          $\mathtt{I_1 ~rel~ I_2 \Leftrightarrow X}$\\
(11) &$\mathtt{int\_op(I_1,I_2,~I)}$ &
      $\mathtt{}$&
      $\mathtt{I_1 ~op~ I_2 = I}$\\
(12) &$\mathtt{int\_array\_op'([I_1,\ldots,I_n],~I)}$ &
      $\mathtt{}$&
      $\mathtt{I_1 ~op'\cdots op'~ I_n = I}$\\
\hline

\multicolumn{4}{l}{\bf\small  All Different and cardinality~
            \hfill $\mathtt{rel{\in}\{leq, geq, eq, lt, gt, neq\}}$}\\
    \hline
(13) &    $\mathtt{allDiff([I_1,\ldots,I_n])}$ &
          &
          $\mathtt{\bigwedge_{i<j}I_i \neq I_j}$\\
(14) &    $\mathtt{bool\_array\_sum\_rel([X_1,\ldots,X_n],~I)}$ &
          $\mathtt{}$&
          $\mathtt{(\Sigma ~X_i)~ rel~ I}$\\
(15) &    $\mathtt{comparator(X_1,X_2,X_3,X_4)}$ &
          &
          $\mathtt{sort([X_1,X_2])=[X_3,X_4]}$\\



\hline\hline
\end{tabular}
  \caption{Syntax for a subset of \bee. }
  \label{tab:beeStntax}
\end{table}

On the right column of the table are brief explanations regarding the
constraints. The table introduces 15 constraint templates.
Constraints (1-2) are about variable declarations: Booleans and
integers. Constraint (3) signifies that a bit sequence is monotonic
non-increasing, and is used to specify that an integer variable is in
the order-encoding.
Constraints (4-7) are about Boolean (and reified Boolean)
statements. The cases for $\mathtt{bool\_array\_or([X_1,\ldots,X_n])}$
and $\mathtt{bool\_array\_xor([X_1,\ldots,X_n])}$ facilitate the
specification of clauses and of \texttt{xor} clauses (supported in the
CryptoMiniSAT solver~\cite{Crypto}).
Constraint (8) specifies that two bit-vectors are ordered
lexicographically. 
Constraints (9-12) are about integer relations and operations.
Constraints (13-14) are the all-different constraint on integers and
the cardinality constraint on Booleans.  Constraint (15) specifies
that sorting a bit pair $\mathtt{[X_1,X_2]}$ (decreasing order)
results in the pair $\mathtt{[X_3,X_4]}$. This is a basic building
block for the construction of sorting networks \cite{Batcher68} used
to encode cardinality constraints during compilation as described
in~\cite{AsinNOR11} and in~\cite{DBLP:conf/lpar/CodishZ10}.

\section{An Example \bee\ Application: magic graph labeling}
\label{sec:magic}

We illustrate the application of \bee\ to solve a
graph labeling problem.
A typical \bee\ application has the form depicted as
Figure~\ref{fig:generic} where the predicate \texttt{solve/2} takes a
problem \texttt{Instance} and provides a \texttt{Solution}. The
specifics of the application are in the call to \texttt{encode/3}
which given the \texttt{Instance} generates the \texttt{Constraints}
that solve it together with a \texttt{Map} relating instance variables with
constraint variables.  The calls to \texttt{compile/2} and 
\texttt{sat/1} compile the constraints to a \texttt{CNF} and solve it
applying a SAT solver. If the instance has a solution, the SAT solver
binds the constraint variables accordingly. Then, the call to
\texttt{decode/2}, using the \texttt{Map}, provides a
\texttt{Solution} in terms of the instance variables.
The definitions of \texttt{encode/3} and \texttt{decode/3} are
application dependent and provided by the user. The predicates
\texttt{compile/2} and \texttt{sat/1} provide the interface to \bee\
and the underlying SAT solver.

\begin{figure}[t]
  \begin{SProg}
    :- use\_module(bee\_compiler, [compile/2]).  \\
    :- use\_module(sat\_solver, [sat/1]).  \\
\vspace{-3mm}\\ 
    solve(Instance, Solution) :- \\
    \qquad    encode(Instance, Map, Constraints),\\
    \qquad    compile(Constraints, CNF), \\
    \qquad    sat(CNF), \\
    \qquad    decode(Map, Solution).\\
  \end{SProg}
  \caption{A generic application of \bee.}
\label{fig:generic}
\end{figure}

Graph labeling is about finding an assignment of integers to the
vertices and edges of a graph subject to certain conditions.  Graph
labelings were introduced in the 60's and hundreds of papers on a wide
variety of related problems have been published since then. See for
example the survey by \citeN{Gallian2011} with more than 1200
references. Graph labelings have many applications. For instance in
radars, xray crystallography, coding theory, etc.

We focus here on the vertex-magic total labeling (VMTL) problem where
one should find for the graph $G=(V,E)$ a labeling that is a
one-to-one map $V\cup E \rightarrow \{1,2,\ldots,|V|+|E|\}$ with the
property that the sum of the labels of a vertex and its incident edges
is a constant $K$ independent of the choice of vertex.
%
A problem instance takes the form $vmtl(G,K)$ specifying the graph $G$
and a constant $K$. The query $\mathtt{solve(vmtl(G,K), Solution)}$
poses the question: ``Does there exist a vmtl labeling for $G$ with
magic constant $K$?'' It binds $\mathtt{Solution}$ to indicate such a
labeling if one exists, or to ``unsat'' otherwise.
\begin{figure}[t]
  \centering
  \begin{tabular}{l}
\hline
\hspace{1cm} An Instance \hspace{2cm} The Graph \hspace{2cm} The Map\\
\hline
$\begin{array}{l}
\mathtt{Instance = vmtl(G,K),}\\
\mathtt{G=(V,E),}\\
\mathtt{V=[1,2,3,4],}\\
\mathtt{E=[(1,2),(1,3),}\\
\qquad   \mathtt{(2,3),(3,4)], }\\
\mathtt{K=14}
\end{array}$\quad\qquad  
$\begin{array}{l}
\xymatrix@C=9pt@R=15pt{ 
            & 4\ar@{-}[d] &            \\
            & 3\ar@{-}[dl]\ar@{-}[dr] &    \\
  2\ar@{-}[rr]&&1 } 
\end{array}$
\qquad\quad$\mathtt{M=} \left[\begin{array}{ll}
   \mathtt{((1,2),~E_1),} & \mathtt{(1,~V_1),} \\
   \mathtt{((1,3),~E_2),} & \mathtt{(2,~V_2),} \\
   \mathtt{((2,3),~E_3),} & \mathtt{(3,~V_3),} \\
   \mathtt{((3,4),~E_4),} & \mathtt{(4,~V_4)} \\
\end{array} \right]$\\
\hline
\hspace{1cm} The Constraints\\
\hline
\qquad$\mathtt{Cs=}
   \left[\begin{array}{lll}
     \mathtt{new\_int(V_{1}, 1, 8),} &
     \mathtt{new\_int(E_{1}, 1, 8),} & 
     \mathtt{int\_array\_plus([V_1,E_1,E_2], K),} 
  \\
     \mathtt{new\_int(V_{2}, 1, 8),} &
     \mathtt{new\_int(E_{2}, 1, 8),} & 
     \mathtt{int\_array\_plus([V_2,E_1,E_3], K),}
  \\
     \mathtt{new\_int(V_{3}, 1, 8),} &
     \mathtt{new\_int(E_{3}, 1, 8),} &
     \mathtt{int\_array\_plus([V_3,E_2,E_3,E_4], K),}
  \\
     \mathtt{new\_int(V_{4}, 1, 8),} &
     \mathtt{new\_int(E_{4}, 1, 8),} & 
     \mathtt{int\_array\_plus([V_4,E_4], K),}
  \\
     \mathtt{new\_int(K, 14, 14),} &
     \multicolumn{2}{l}{
     \mathtt{allDiff([V_1,V_2,V_3,V_4,E_1,E_2,E_3,E_4])}} 
 \end{array}\right]$\\
\hline
\end{tabular}

  \caption{A VMTL instance with the constraints and map generated by
    \texttt{encode/3}. }
  \label{fig:vmtl-instance}
\end{figure}
Figure~\ref{fig:vmtl-instance} illustrates an example problem instance
together with the constraints, \texttt{Cs} and the map, \texttt{M},
generated by the \texttt{encode/3} predicate for this instance.
The constraints introduce integer variables for the vertices and
edges, specify that these variables take  ``all different'' values,
and specify that the labels for each vertex with its incident edges
sum to $\mathtt{K}$.
Solving the constraints from Figure~\ref{fig:vmtl-instance} for the
example VMTL instance binds the Map, \texttt{M}, as follows,
indicating a solution:
\[\mathtt{M=}{\small \left[\begin{array}{ll}
\mathtt{((1,2),~[1,1,1,1,1,1,1,0]),} & \mathtt{(1,~[1,1,1,1,0,0,0,0]),} \\
\mathtt{((1,3),~[1,1,1,0,0,0,0,0]),} & \mathtt{(2,~[1,1,1,1,1,0,0,0]),} \\
\mathtt{((2,3),~[1,1,0,0,0,0,0,0]),} & \mathtt{(3,~[1,0,0,0,0,0,0,0]),} \\
\mathtt{((3,4),~[1,1,1,1,1,1,1,1]),} & \mathtt{(4,~[1,1,1,1,1,1,0,0])} \\
\end{array} \right]}
\]
In Section~\ref{results} we report that using \bee\ enables
us to solve interesting instances of the VMTL problem not previously
solvable by other techniques.

\section{Compiling \bee\ to CNF}
\label{sec:compiling}

The compilation of a constraint model to a CNF
using \bee\ goes through three phases. 
In the first phase, (unary) bit blasting, integer variables (and
constants) are represented as bit vectors in the order-encoding. Now
all constraints are about Boolean variables.
The second phase, the main loop of the compiler, is about constraint
simplification.  Three types of actions are applied: equi-propagation,
partial evaluation, and decomposition of constraints. These are
specified as a set of transitions which we write in the form
$c_1\overset{\theta}{\longmapsto}c_2$ to specify that constraint $c_1$
reduces to constraint $c_2$ generating the (possibly empty)
substitution $\theta$. Simplification is applied repeatedly until no
rule is applicable.
In the third, and final phase, simplified constraints are encoded to
CNF. We elaborate below.
To simplify the presentation, we assume that integer variables are
represented in a positive interval starting from~$0$. As later
detailed in Section~\ref{sec:implementation} there is no such
limitation in \bee.

\vspace{-3mm}
\paragraph{\underline{Bit-blasting:}} 

Each integer variable declaration
$\mathtt{new\_int(I,c_1,c_2)}$ triggers a unification
$\mathtt{I=[1,\dots,1,X_{c_1+1},\ldots,X_{c_2}]}$ and introduces a
constraint $\mathtt{ordered(I)}$ to specify that the bits representing
$\mathtt{I}$ are in the order-encoding.  
To illustrate bit-blasting, consider again the VMTL
example detailed in Figure~\ref{fig:vmtl-instance}. Each variable in
the \texttt{Map} occurs in a $\mathtt{new\_int}$ declaration. So the
following unifications are performed:
\[{\small\begin{array}{ll}
\mathtt{V_{1}=[1,V_{1,2},V_{1,3},V_{1,4},V_{1,5},V_{1,6},V_{1,7},V_{1,8}]}, &
\mathtt{E_{1}=[1,E_{1,2},E_{1,3},E_{1,4},E_{1,5},E_{1,6},E_{1,7},E_{1,8}]}, \\
\mathtt{V_{2}=[1,V_{2,2},V_{2,3},V_{2,4},V_{2,5},V_{2,6},V_{2,7},V_{2,8}]}, &
\mathtt{E_{2}=[1,E_{2,2},E_{2,3},E_{2,4},E_{2,5},E_{2,6},E_{2,7},E_{2,8}]}, \\
\mathtt{V_{3}=[1,V_{3,2},V_{3,3},V_{3,4},V_{3,5},V_{3,6},V_{3,7},V_{3,8}]}, &
\mathtt{E_{3}=[1,E_{3,2},E_{3,3},E_{3,4},E_{3,5},E_{3,6},E_{3,7},E_{3,8}]}, \\
\mathtt{V_{4}=[1,V_{4,2},V_{4,3},V_{4,4},V_{4,5},V_{4,6},V_{4,7},V_{4,8}]}, & 
\mathtt{E_{4}=[1,E_{4,2},E_{4,3},E_{4,4},E_{4,5},E_{4,6},E_{4,7},E_{4,8}]},\\
\mathtt{K=~[1,1,1,1,1,1,1,1,1,1,1,1,1,1]}&
\end{array}}\]
Integer variables occurring in an \texttt{allDiff} constraint are
bit-blasted twice: first, in the order-encoding, when declared, as
explained above, and second, in the direct encoding, when processing
the \texttt{allDiff} constraint, as described below.

\vspace{-3mm}
\paragraph{\underline{Equi-propagation}} is about detecting
situations in which a small number of constraints imply an equality of
the form $X=L$ where $X$ is a Boolean variable and $L$ is a Boolean
literal or constant. In this case $X$ becomes redundant
and can be replaced by $L$ in all constraints.
In \bee\ we consider as candidates for equi-propagation, individual
constraints together with constraints specifying that their integer
variables are in the order-encoding. If $X=L$ is such an equality,
then equi-propagation is implemented by unifying $X$ and $L$. This
unification applies to all occurrences of $X$ and in this sense
``propagates'' to other constraints involving $X$. Once
equi-propagation detects such an equation, this may trigger further
equi-propagation from other constraints.
For example, consider the constraint $\mathtt{int\_neq(I_1,I_2)}$
where $\mathtt{I_1=[x_1,x_2,x_3,x_4}]$ and
$\mathtt{I_2=[1,1,0,0}]$. We propagate that $\mathtt{(x_2=x_3)}$ because
\[
\left(
  \begin{array}{l}
    \mathtt{I_1=[x_1,x_2,x_3,x_4] ~\wedge~ I_2=[1,1,0,0] ~\wedge}\\
    \mathtt{int\_neq(I_1,I_2) ~\wedge~ ordered(I_1)}
  \end{array}
\right)\models \mathtt{(x_2=x_3)}.
\]
To see why, consider that $\mathtt{ordered(I_1)}$ implies that
$\mathtt{x_2\geq x_3}$. Furthermore, also $\mathtt{x_2\leq x_3}$ as
otherwise $\mathtt{x_2=1}$ and $\mathtt{x_3=0}$ which implies that
$\mathtt{I_1=[1,1,0,0}]$, contradicting $\mathtt{int\_neq(I_1,I_2)}$.

In \bee, equi-propagation is implemented by a collection of ad-hoc
transition rules for each type of constraint. While this approach is
not complete --- there are equations implied by a constraint that
\bee\ will not detect --- the implementation is fast, and works well
in practice.
An alternative approach is to implement equi-propagation, using BDD's,
as described in~\cite{Metodi2011}. This approach, though complete, is
slower and not included in the current release of \bee.

The following are two of the simplification (equi-propagation) rules
of \bee\ that apply to $\mathtt{int\_neq}$ constraints:

\begin{description}
\item[$\mathtt{neq_1}:$] applies when one of the (order-encoding)
  integers in the relation is a constant and $\theta=\{X_1=X_2\}$:
\[
\mathtt{int\_neq}\left( 
\begin{array}{cccc}
{[}\ldots,\hspace{-3mm}&X_1,\hspace{-3mm}&X_2,\hspace{-3mm}& \ldots{]} \\
{[}\ldots,\hspace{-3mm}&1,  \hspace{-3mm}& 0,\hspace{-3mm}&  \ldots{]}
\end{array}
\right)
\overset{\theta}{\longmapsto}
\mathtt{int\_neq}\left(
\begin{array}{cccc}
{[}\ldots,\hspace{-3mm}&X_1,\hspace{-3mm}&X_1,\hspace{-3mm}& \ldots{]} \\
{[}\ldots,\hspace{-3mm}&1,  \hspace{-3mm}& 0,\hspace{-3mm}&  \ldots{]}
\end{array}
\right)
\]

\item[$\mathtt{neq_2}:$] applies when the integers share common
  variables as in the rule template and $\theta=\{X_1=X_2\}$:
\[
\mathtt{int\_neq}\left( 
\begin{array}{cccc}
{[}\ldots,\hspace{-3mm}&X_1,\hspace{-3mm}&X_2,\hspace{-3mm}&\ldots{]}, \\
{[}\ldots,\hspace{-3mm}&\neg X_2,\hspace{-3mm}&\neg X_1,\hspace{-3mm}&\ldots{]}
\end{array}
\right)
\overset{\theta}{\longmapsto}
\mathtt{int\_neq}\left(
\begin{array}{cccc}
{[}\ldots,\hspace{-3mm}&X_1,\hspace{-3mm}&X_1,\hspace{-3mm}&\ldots{]}, \\
{[}\ldots,\hspace{-3mm}&\neg X_1,\hspace{-3mm}&\neg X_1,\hspace{-3mm}&\ldots{]}
\end{array}
\right)
\]
\end{description}

For the  rule $\mathtt{neq_1}$, observe that after applying this
rule the constraint obtained is a tautology. Hence it is subsequently
removed by one of the other ``partial evaluation'' rules.
For the rule $\mathtt{neq_2}$, to see why the equation
$X_1=X_2$ is implied by the constraint (on the left side of the rule),
consider all possible truth values for the variables $X_1$ and $X_2$:
(a) If $X_1=0$ and $X_2=1$ then both integers in the relation take the
form $[\ldots,0,1,\ldots]$ violating their specification as
\texttt{ordered}, so this is not possible. (b) If $X_1=1$ and $X_2=0$
then both numbers take the form $[1,\ldots,1,0,\ldots,0]$ and are
equal, violating the $\mathtt{neq}$ constraint. The only possible
bindings for $X_1$ and $X_2$ are those where $X_1=X_2$. 
The template expressed in rule $\mathtt{neq_2}$ is not contrived. It
comes up frequently as a result of applying other equi-propagation
rules. 

\vspace{-3mm}
\paragraph{\underline{Partial evaluation}} is about simplifying
constraints in view of variables that are (partially) instantiated,
either because of information from the constraint model or else due to
equi-propagation. Typical cases include constant elimination and
elimination of tautologies.
The following are some of \bee's partial evaluation rules that apply
to $\mathtt{int\_neq}$ constraints ($\epsilon$ denotes the empty
substitution). 

\begin{description}
\item[$\mathtt{neq_3}:$] applies to remove replicated variables:
\[
\mathtt{int\_neq}\left( 
\begin{array}{cccc}
{[}\ldots,\hspace{-3mm}&X_1,\hspace{-3mm}&X_1,\hspace{-3mm}& \ldots{]} \\
{[}\ldots,\hspace{-3mm}&Y_1,\hspace{-3mm}&Y_1,\hspace{-3mm}&  \ldots{]}
\end{array}
\right)
\overset{\epsilon}{\longmapsto}
\mathtt{int\_neq}\left(
\begin{array}{ccc}
{[}\ldots,\hspace{-3mm}&X_1,\hspace{-3mm}& \ldots{]} \\
{[}\ldots,\hspace{-3mm}&Y_1,\hspace{-3mm}&  \ldots{]}
\end{array}
\right)
\]

\item[$\mathtt{neq_4}:$] applies to remove leading 1 bits (there is a
  similar rule for trailing 0's):
\[
\mathtt{int\_neq}([1,1,X_3,\ldots], [Y_1,Y_2,Y_3,\ldots]) 
   \overset{\epsilon}{\longmapsto}
 \mathtt{int\_neq}([1,X_3,\ldots],[Y_2,Y_3,\ldots])
\]
\end{description}

We now detail three of the simplification rules (equi-propagation and
partial evaluation) that apply to a constraint of the form
\texttt{int\_plus(A,B,C)} where we assume for simplicity of
presentation (the tool supports the general case) that
$\mathtt{A=[A_1,\ldots,A_n]}$, ~$\mathtt{B=[B_1,\ldots,B_m]}$, and
$\mathtt{C=[C_1,\ldots,C_{n+m}]}$. We denote by $\mathtt{min(I)}$ (or
$\mathtt{max(I)}$) the minimal (or maximal) value that integer
variable \texttt{I} can take, determined by the number of leading ones
(or trailing zeros) in its bit representation.

Rule $\mathtt{plus_{1}}$ is standard propagation for interval
arithmetics.  Rule $\mathtt{plus_2}$ removes redundant bits 
(assigned values through $\mathtt{plus_{1}}$).
Rules $\mathtt{plus_{3(a)}}$ and $\mathtt{plus_{3(b)}}$ remove
constraints and may seem contrived: \texttt{{3(a)}} assumes that
$\mathtt{m=0}$ and \texttt{{3(b)}} assumes that $\mathtt{n=m}$ and
that $\mathtt{C}$ represents the (same) constant $\mathtt{n}$. 
However, in the general case, when $\mathtt{n}$, $\mathtt{m}$ are
arbitrary and constant \texttt{C} is represented in $\mathtt{m+n}$
bits, then application of the other rules will reduce the constraint
to one of these special cases.

\begin{description}
\item[$\mathtt{plus_1}:$] applies to propagate bounds:
$
\mathtt{int\_plus(A,B,C)} 
   \overset{\theta}{\longmapsto}
\mathtt{int\_plus(A,B,C)} 
$
where \\
\hspace*{-7mm}$\theta=\left\{
\begin{array}{ll}
C_{max\{min(C), min(A)+min(B)\}} = 1, & C_{min\{max(C), max(A)+max(B)\}+1} = 0,\\
A_{max\{min(A), min(C)-max(B)\}} = 1, & A_{min\{max(A), max(C)-min(B)\}+1} = 0, \\
B_{max\{min(B), min(C)-max(A)\}} = 1, & B_{min\{max(B), max(C)-min(A)\}+1} = 0
\end{array}\right\}
$\smallskip
\item[$\mathtt{plus_{2}}:$] applies to remove leading 1's  
 (there are similar rules for trailing 0's and for the case when
 the 1's or 0's are on $\mathtt{[B_1,\ldots,B_m]}$):
\[
\mathtt{int\_plus}\left(\begin{array}{l}
  \mathtt{[1,A_2,\ldots,A_{n}], }\\
  \mathtt{[B_1,\ldots,B_m],}\\
  \mathtt{{[}1,C_2,\ldots,C_{n+m}{]}} 
\end{array}\right)
   \overset{\epsilon}{\longmapsto}
\mathtt{int\_plus}\left(\begin{array}{l}
  \mathtt{[A_2,\ldots,A_{n}], }\\
  \mathtt{[B_1,\ldots,B_m],}\\
  \mathtt{{[}C_2,\ldots,C_{n+m}{]}} 
\end{array}\right)
\]
\item[$\mathtt{plus_{3(a)}}:$] applies when \texttt{A} or \texttt{B}
  is the empty bit list and $\theta = \sset{C_{i}=A_i}{1 \leq i \leq
    n}$
\[
\mathtt{int\_plus([A_1,\ldots,A_n], [~], [C_1,\ldots,C_n])}
   \overset{\theta}{\longmapsto} none
\]

\item[$\mathtt{plus_{3(b)}}:$] applies when \texttt{C} is a constant
  \texttt{n} and $\theta = \sset{A_{i}=\neg B_{n-i+1}}{1 \leq i
    \leq n}$
\[
\mathtt{int\_plus([A_1,\ldots,A_n], [B_1,\ldots,B_n], [1,\ldots1,0,\ldots,0])}
   \overset{\theta}{\longmapsto}
none
\]
  
\end{description}

We illustrate the simplification of a \texttt{int\_plus} constraint by
the following example.

\begin{example}[simplifying \texttt{int\_plus}: equi-propagation and
  partial evaluation]\label{ex:plus14}
  Consider constraint $\mathtt{int\_plus(A,B,C)}$ where $A$ and $B$
  are integer variables with domain $[1..8]$ and $C$ is the constant
  14 represented in 16 bits. Constraint simplification follows the
  steps:

\medskip
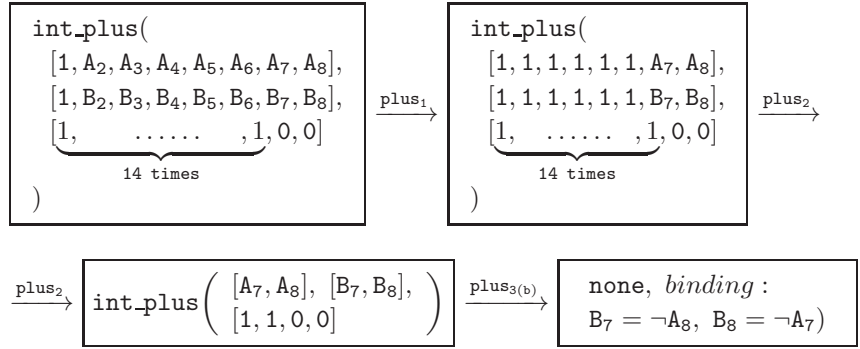
\begin{tabular}{l}
\fbox{$\begin{array}{l}
\mathtt{int\_plus(}\\
~~\mathtt{[1,A_2,A_3,A_4,A_5,A_6,A_7,A_8],} \\
~~\mathtt{[1,B_2,B_3,B_4,B_5,B_6,B_7,B_8],}\\
~~\mathtt{[\underbrace{1,\qquad\ldots\ldots~\quad,1}_{14
    ~times},0,0]}\\
\mathtt{)}
\end{array}$} 
$\xrightarrow{\mathtt{plus_1}}$
\fbox{$\begin{array}{l}
\mathtt{int\_plus(}\\
~~\mathtt{[1,1,1,1,1,1,A_7,A_8],} \\
~~\mathtt{[1,1,1,1,1,1,B_7,B_8],}\\
~~\mathtt{[\underbrace{1,\quad\ldots\ldots~~,1}_{14 ~times},0,0]}\\
\mathtt{)}
\end{array}$}
$\xrightarrow{\mathtt{plus_{2}}}$ 
\\ ~\\
$\xrightarrow{\mathtt{plus_{2}}}$
\fbox{\texttt{int\_plus}$\left(\begin{array}{l}
\mathtt{[A_7,A_8],~} 
\mathtt{[B_7,B_8],}\\
\mathtt{[1,1,0,0]}
\end{array}\right)$}
$\xrightarrow{\mathtt{plus_{3(b)}}}$
\fbox{
  $\begin{array}{l}
    \mathtt{none}, ~binding:  \\
     \mathtt{B_7=\neg A_8,~B_8=\neg A_7})
  \end{array}$

} 
\end{tabular}

\medskip\noindent After constraint simplification
variables \texttt{A} and \texttt{B} take the form: $\mathtt{[1, 1, 1,
  1, 1, 1, A_{7}, A_{8}]}$ and $\mathtt{[1, 1, 1, 1, 1,
  1,\neg A_{8}, \neg A_{7}]}$ (and nothing is left to encode to CNF).

\end{example}

\vspace{-3mm}
\paragraph{\underline{Decomposition}} is about replacing complex
constraints (for example about arrays) with simpler constraints (for
example about array elements). Consider, for instance, the constraint
$\mathtt{int\_array\_plus(As,Sum)}$. It is decomposed to a list of
$\mathtt{int\_plus}$ constraints applying a straightforward divide and
conquer recursive definition. At the base case, if \texttt{As=[A]}
then the constraint is replaced by \texttt{int\_eq(A,Sum)}, or if
$\mathtt{As=[A_1,A_2]}$ then it is replaced by
$\mathtt{int\_plus(A_1,A_2,Sum)}$.
In the general case \texttt{As} is split into two halves, then
constraints are generated to sum these halves, and then an additional
$\mathtt{int\_plus}$ constraint is introduced to sum the two sums.
%




As another example, consider the $\mathtt{int\_plus(A_1,A_2,A)}$
constraint.  One approach, supported by \bee, decomposes the
constraint as an odd-even merger (from the context of odd-even sorting
networks)~\cite{Batcher68}. Here, the sorted sequences of bits
$\mathtt{A_1}$ and $\mathtt{A_2}$ are merged to obtain their sum
$\mathtt{A}$.
This results in a model  with $O(n\log n)$
\texttt{comparator} constraints (and later in an encoding with
$O(n\log n)$ clauses).
Another approach, also supported in \bee, does not decompose the
constraint but encodes it directly to a CNF of size $O(n^2)$,
as in the context of so-called totalizers~\cite{BailleuxB03}.
A hybrid approach, leaves the choice to \bee, depending on the size of
the domains of the variables involved. 
Finally, we note that the user can configure \bee\ to fix the way it
compiles this constraint (and others).






\paragraph{\underline{CNF encoding}} is the last phase and applies to
all remaining simplified constraints. The encoding of constraints to
CNF is standard and similar to the encodings in
Sugar~\cite{sugar2009}.

\paragraph{\underline{Cardinality constraints}} are about the
cardinality of sets of Boolean variables and are specified by the
template $\mathtt{bool\_array\_sum\_rel([X_1,\ldots,X_n],~I)}$.
Cardinality constraints are normalized, see e.g., \cite{EenS06}, so we
only consider $\mathtt{rel\in\{leq,eq\}}$. Partial evaluation rules
for cardinality constraints are the obvious. For example, in the
special case when \texttt{I} is a constant:

$\mathtt{bool\_array\_sum\_leq([X_1,X_2,1,X_4],~3)\mapsto
  bool\_array\_sum\_leq([X_1,X_2,X_4],~2)}$

$\mathtt{bool\_array\_sum\_leq([X_1,X_2,0,X_4],~3)\mapsto
  bool\_array\_sum\_leq([X_1,X_2,X_4],~3)}$

$\mathtt{bool\_array\_sum\_leq([X_1,X_2,-X_1,X_4],~3)\mapsto
  bool\_array\_sum\_leq([X_2,X_4],~2)}$

\medskip\noindent 
The special case, when \texttt{I} is the constant~1 is called the
``at-most-one'' constraint and it has been studied extensively (for a
recent survey see \cite{Frisch+Giannaros/10/SAT}). In \bee, we support
two different encodings for this case (the user can choose). The first
is the standard ``pairwise'' encoding which specifies a clause $(\neg
x_i\vee\neg x_j)$ for each pair of Boolean variables $x_i$ and $x_j$.
This encoding introduces $O(n^2)$ clauses and is sometimes too large.
The second,  is a more compact encoding which follows
the approach described in~\cite{NewAtMostOne}.
In the general case (when $\mathtt{I}>1$) the constraint is
decomposed, much the same as the $\mathtt{int\_array\_plus}$
constraint, to a network of $\mathtt{int\_plus}$ constraints.

\paragraph{\underline{The All-different constraint}}
specifies that a set of integer variables take all different
values. Although we adopt the order-encoding for integer variables, it
is well accepted that for these constraints the direct encoding is
superior \cite{direct4allDiff}. For this reason, in \bee, when
processing the constraint, a dual representation is chosen.  When
integer variable $\mathtt{I}$, occurring in an \texttt{allDiff}
constraint, is declared, it was unified with its unary representation
in the order-encoding: $\mathtt{I=[x_1,\ldots,x_n]}$. In addition, we
associate \texttt{I} with a new bit-blast,
$\mathtt{[d_0,\ldots,d_{n}]}$, in the direct encoding.  We introduce
for each such \texttt{I} a channeling formula to capture the relation
between its two representations.
\[\mathtt{channel([x_1,\ldots,x_n],[d_0,\ldots,d_{n}])=
  \left(\begin{array}{r} d_0 = \neg x_1 \\
                         \wedge~ d_n = x_n
  \end{array}\right) \wedge
  \bigwedge_{i=1}^{n-1}(d_i\leftrightarrow x_{i}\wedge\neg x_{i+1})}
\]

During constraint simplification, the
$\mathtt{allDiff([I_1,\ldots,I_n])}$ constraint is viewed as a bit
matrix where each row consists of the bits
$\mathtt{[d_{i0},\ldots,d_{im}]}$ for $\mathtt{I_i}$ in the direct
encoding. The element $d_{ij}$ is true iff $I_i$ takes the value $j$.
The $j^{th}$ column specifies which of the $I_i$ take the value $j$
and hence, at most one variable in a column may take the value true.
\bee\ distinguishes the special case when $\mathtt{[I_1,\ldots,I_n]}$
must take precisely $n$ different values. In this case the constraint
is about ``permutation''. We denote this by a flag (*) as in
$\mathtt{allDiff^*([I_1,\ldots,I_n])}$. In this case, exactly one bit
in each column of the representation must take the value true.

To simplify an \texttt{allDiff} constraint, \bee\ applies
simplification rules to the implicit cardinality constraints on the
columns and also two specific \texttt{allDiff} rules. The first is
essentially the usual domain consistent propagator~\cite{regin}
focusing on Hall sets of size 2. The second rule applies only to an
$\mathtt{allDiff^*}$ constraint which is about permutation.
We denote the values that $\mathtt{I_i}$ can take as
$\mathtt{dom(I_i)}$.

\begin{description}
\item[$\mathtt{allDiff_1}:$]  
 when $\mathtt{dom(I_1) = dom(I_2)=\{v_1,v_2\}}$:
\[
\mathtt{allDiff([I_1,I_2,I_3,\ldots,I_n])}
\overset{\theta}{\longmapsto}
\mathtt{allDiff([I_3,\ldots,I_n])}
\]
  where $\theta = \bigcup_{3\leq i\leq n}
                    \{d_{i,v_1}=0, d_{i,v_2}=0\} \cup
                 \{d_{1,v_1}= -d_{2,v_1}, d_{1,v_2}= -d_{2,v_2}  \}$.

\item[$\mathtt{allDiff_2}:$]  when
  $\mathtt{\{v_1,v_2\}\subseteq dom(I_1)\cap dom(I_2)}$,
   and for $i\geq 3$,
  $\mathtt{\{v_1,v_2\}\cap dom(I_i)=\emptyset}$ 
\[
\mathtt{allDiff^*([I_1,\ldots,I_n])}
\overset{\theta}{\longmapsto}
\mathtt{allDiff^*([I_1,\ldots,I_n])}
\]
  where $\theta = \bigcup_{j\neq v_1, j\neq v_2}\{d_{1j}=0, d_{2,j}=0\}$.

\end{description}

To illustrate the two rules for \texttt{allDiff} consider the
following.  
\begin{example}
  Consider an \texttt{allDiff} constraint on 5 integer variables
  taking values in the interval $[0,7]$ where the first two can take
  only values 0 and 1. So, they are a Hall set of size two and rule
  $\mathtt{allDiff_1}$ applies. We present the simplification step on
  the order encoding representation (though it is triggered through
  the direct encoding representation): \vspace{-2mm}
\[\small
\mathtt{allDiff}\left( 
\begin{array}{cccc}
{[}X_{1,1},\hspace{-3mm}&0,\hspace{-3mm}& \ldots,\hspace{-3mm}&0{]} \\
{[}X_{2,1},\hspace{-3mm}&0,\hspace{-3mm}& \ldots,\hspace{-3mm}&0{]} \\
{[}X_{3,1},\hspace{-3mm}&X_{3,2},\hspace{-3mm}& \ldots,\hspace{-3mm}&X_{3,7}{]} \\
{[}X_{4,1},\hspace{-3mm}&X_{4,2},\hspace{-3mm}& \ldots,\hspace{-3mm}&X_{4,7}{]} \\
{[}X_{5,1},\hspace{-3mm}&X_{5,2},\hspace{-3mm}& \ldots,\hspace{-3mm}&X_{5,7}{]}
\end{array}
\right)
\overset{\theta}{\longmapsto}
\mathtt{allDiff}\left(
\begin{array}{ccc}
{[}1,X_{3,2},\hspace{-3mm}& \ldots,\hspace{-3mm}&X_{3,7}{]} \\
{[}1,X_{4,2},\hspace{-3mm}& \ldots,\hspace{-3mm}&X_{4,7}{]} \\
{[}1,X_{5,2},\hspace{-3mm}& \ldots,\hspace{-3mm}&X_{5,7}{]}
\end{array}
\right)
\]
where $\theta = \{ X_{1,1}=\neg X_{2,1} , X_{3,1}=0 , X_{4,1}=0 ,
X_{5,1}=0\}$.

Now consider a setting where an \texttt{allDiff} constraint is about 5
variables that can take 5 values (permutation) and the first two are
the only two that can take values 0 and 1. So rule
$\mathtt{allDiff_2}$ applies. We present the simplification step on
  the order encoding representation (though it is triggered through
  the direct encoding representation):
\vspace{-2mm}
\[\small
\mathtt{allDiff^*}\left( 
\begin{array}{lccr}
{[}X_{1,1},\hspace{-3mm}&X_{1,2},\hspace{-3mm}&X_{1,3},\hspace{-3mm}&X_{1,4}{]} \\
{[}X_{2,1},\hspace{-3mm}&X_{2,2},\hspace{-3mm}&X_{2,3},\hspace{-3mm}&X_{2,4}{]} \\
{[}1,\hspace{-3mm}&1,\hspace{-3mm}&X_{3,3},\hspace{-3mm}&X_{3,4}{]} \\
{[}1,\hspace{-3mm}&1,\hspace{-3mm}&X_{4,3},\hspace{-3mm}&X_{4,4}{]} \\
{[}1,\hspace{-3mm}&1,\hspace{-3mm}&X_{5,3},\hspace{-3mm}&X_{5,4}{]}
\end{array}
\right)
\overset{\theta}{\longmapsto}
\mathtt{allDiff^*}\left(
\begin{array}{lccr}
{[}X_{1,1},\hspace{-3mm}&0,\hspace{-3mm}&0,\hspace{-3mm}&0{]} \\
{[}X_{2,1},\hspace{-3mm}&0,\hspace{-3mm}&0,\hspace{-3mm}&0{]} \\
{[}1,\hspace{-0mm}&1,\hspace{-0mm}&X_{3,3},\hspace{-3mm}&X_{3,4}{]} \\
{[}1,\hspace{-0mm}&1,\hspace{-0mm}&X_{4,3},\hspace{-3mm}&X_{4,4}{]} \\
{[}1,\hspace{-0mm}&1,\hspace{-0mm}&X_{5,3},\hspace{-3mm}&X_{5,4}{]}
\end{array}
\right)
\]
  where $\theta = \{ X_{1,2},\ldots,X_{1,4} = 0 , X_{2,2},\ldots,X_{2,4} = 0\}$.

\end{example}

When no further simplification rules apply the \texttt{allDiff}
constraint is decomposed to the corresponding cardinality constraints
on the columns of its bit matrix representation.

\section{Constraint simplification in the VMTL example}

Consider again the VMTL example and the constraints from
Figure~\ref{fig:vmtl-instance}. We focus on three
constraints and follow the steps made when compiling these (we write
``14'' as short for $\mathtt{[\underbrace{1,1,\ldots,1}_{14}]}$). 

\smallskip
$\begin{array}{ll}
(1) & \mathtt{int\_array\_plus([V_4,E_4], 14)} \\
(2) & \mathtt{allDiff([V_1,V_2,V_3,V_4,E_1,E_2,E_3,E_4]),} \\
(3) & \mathtt{int\_array\_plus([V_3,E_2,E_3,E_4], 14),} 
\end{array}$

\smallskip\noindent In the first steps, constraint (1) is decomposed to
an \texttt{int\_plus} constraint which has the same form as the
constraint in Example~\ref{ex:plus14}. So, we have the bindings
$\mathtt{V_4 = [1,1,1,1,1,1,V_{4,7}, V_{4,8}]}$ and
$\mathtt{E_4 = [1,1,1,1,1,1,\neg V_{4,8},\neg V_{4,7}]}$. 
Now, consider the \texttt{allDiff} constraint (2). \bee\ determines
that this constraint is about permutation (8 integer variables with 8
different values in the range [1,8]). The simplification rules for
\texttt{allDiff} detect that $\mathtt{\{V_4,E_4\}}$ must take together
the two values 6 and 8 (using a simplification rule similar to
$\mathtt{neq_2}$) triggerring the substitution $\mathtt{\{V_{4,7} =
  V_{4,8}\}}$.  Now rule $\mathtt{allDiff_1}$ detects a Hall set
$\mathtt{\{V_4,E_4\}}$ of size two:
\[\small \mathtt{allDiff([V_1,V_2,V_3,V_4,E_1,E_2,E_3,E_4])}
\xrightarrow{\theta} 
\mathtt{allDiff([V_1,V_2,V_3,E_1,E_2,E_3])}
\]
where $\theta$ is the unification that imposes
$\mathtt{V_1,V_2,V_3,E_1,E_2,E_3 \neq 6,8}$. So we have the following
bindings (where the impact of $\theta$ is underlined): 
\[\small \begin{array}{ll}
\mathtt{V_{1}=
   [1,V_{1,2},V_{1,3},V_{1,4},V_{1,5},\underline{V_{1,7},V_{1,7}},0]} \qquad&
\mathtt{E_{1}=
   [1,E_{1,2},E_{1,3},E_{1,4},E_{1,5},\underline{E_{1,7},E_{1,7}},0]} \\
\mathtt{V_{2}=
   [1,V_{2,2},V_{2,3},V_{2,4},V_{2,5},\underline{V_{2,7},V_{2,7}},0]} &
\mathtt{E_{2}=
   [1,E_{2,2},E_{2,3},E_{2,4},E_{2,5},\underline{E_{2,7},E_{2,7}},0]} \\
\mathtt{V_{3}=
   [1,V_{3,2},V_{3,3},V_{3,4},V_{3,5},\underline{V_{3,7},V_{3,7}},0]} &
\mathtt{E_{3}=
   [1,E_{3,2},E_{3,3},E_{3,4},E_{3,5},\underline{E_{3,7},E_{3,7}},0]} \\
\mathtt{V_{4}=
   [1,1,1,1,1,1,V_{4,7},V_{4,7}]} & 
\mathtt{E_{4}=
   [1,1,1,1,1,1,\neg V_{4,7},\neg V_{4,7}]}
\end{array}  
\]




Consider now the constraint (3). Equi-propagation (because of bounds)
dictates that $\mathtt{max(V_1)=max(V_2)=max(V_3)=5}$, so this
constraint then simplifies as follows:

\[\small
   \fbox{$\begin{array}{l}
    \mathtt{int\_array\_plus([}\\
    ~~\mathtt{[1,V_{3,2},V_{3,3},V_{3,4},V_{3,5},0,0,0]},\\
    ~~\mathtt{[1,E_{2,2},E_{2,3},E_{2,4},E_{2,5},0,0,0]},\\
    ~~\mathtt{[1,E_{3,2},E_{3,3},E_{3,4},E_{3,5},0,0,0]},\\
    ~~\mathtt{[1,1,1,1,1,1,\neg V_{4,7},\neg V_{4,7}]},~14~
   \mathtt{])}
  \end{array}$}
\longmapsto
   \fbox{$\begin{array}{l}
    \mathtt{int\_array\_plus([}\\
    ~~\mathtt{[V_{3,2},V_{3,3},V_{3,4},V_{3,5}]},\\
    ~~\mathtt{[E_{2,2},E_{2,3},E_{2,4},E_{2,5}]},\\
    ~~\mathtt{[E_{3,2},E_{3,3},E_{3,4},E_{3,5}]},\\
    ~~\mathtt{[\neg V_{4,7},\neg V_{4,7}]},~5~
    \mathtt{])}
  \end{array}$}
\]

\medskip After applying simplification and decomposition rules on all
the constraints from Figure~\ref{fig:vmtl-instance} until no further
rules can be applyed, the constraints will be encoded to CNF.  The
generated CNF contains 301 clauses and 48 Boolean variables.
Compiling the same set of constraints from
Figure~\ref{fig:vmtl-instance} without applying simplification rules
generates a larger CNF which contains 642 clauses and 97 Boolean
variables.



\section{Another Example \bee\ Application: 
DNA word problem}

The DNA word problem (Problem \texttt{033} of CSPLib) seeks the
largest parameter $n$, such that there exists a set $S$ of $n$
eight-letter words over the alphabet $\Sigma=\{A,C,G,T\}$ with the
following properties:
\textbf{(1)} Each word in $S$ has exactly 4 symbols from $\{C,G\}$;
\textbf{(2)} Each pair of distinct words in $S$ differ in at least 4 positions;
and
\textbf{(3)} For every $x,y\in S$: $x^R$ (the reverse of $x$) and $y^C$ (the
word obtained by replacing each $A$ by $T$, each $C$ by $G$, and vice
versa) differ in at least 4 positions.

In~\cite{dnaWordPaper}, the authors present a strategy to solve this
problem where the four letters are modeled by bit-pairs
$\tuple{t,m}$. Each eight-letter word can then be viewed as the
combination of a \emph{``t-part''}, $\tuple{t_1,\ldots,t_8}$, which is
a bit-vector, and a \emph{``m-part''}, $\tuple{m_1,\ldots,m_8}$, also
a bit-vector. Building on the approach described
in~\cite{dnaWordPaper}, we pose conditions on sets of
\emph{``t-parts''} and \emph{``m-parts''}, $T$ and $M$, so that their
Cartesian product $S=T\times M$ will satisfy the requirements of the
original problem. From the three conditions below, $T$ is required to
satisfy (1$'$) and (2$'$), and $M$ is required to satisfy (2$'$) and
(3$'$).
For a set of bit-vectors $V$, the conditions are:
\textbf{(1$'$)} Each bit-vector in $V$ sums to 4;
\textbf{(2$'$)} Each pair of distinct bit-vectors in $V$ differ in at
  least 4 positions; and
  \textbf{(3$'$}) For each pair of bit-vectors (not necessarily distinct)
  $u,v\in V$, $u^R$ (the reverse of $u$) and $v^C$ (the complement of
  $v$) differ in at least 4 positions. This is equivalent to requiring
  that $(u^r)^c$ differs from $v$ in at least 4 positions.

  It is this strategy that we model in our \bee\ encoding.
  An instance takes the form $\mathtt{dna(n_1,n_2)}$ signifying the
  numbers of bit-vectors, $n_1$ and $n_2$ in the sets $T$ and $M$.
  Without loss of generality, we impose, to remove symmetries, that
  $T$ and $M$ are lexicographically ordered.
  A solution is the Cartesian product $S=T\times M$.
  In Section~\ref{results} we report that using \bee\ enables us to
  solve interesting instances of the problem not previously solvable
  by other techniques.

\section{Implementation}
\label{sec:implementation}

\bee\ is implemented in (SWI) Prolog and can be applied in conjunction
with the CryptoMiniSAT solver \cite{Crypto} through a Prolog interface
\cite{satPearl}. \bee\ can be downloaded from \cite{bee2012} where one
can find also the examples from this paper and others. The
distribution includes also a solver, which we call Bumble\bee, which
enables to specify a \bee\ model as an input file and solve it. The
output is a set of bindings to the declared variables in the model.

In \bee, Boolean variables are represented as Prolog variables. The
negation of \texttt{X} is represented as \texttt{-X}. The truth
values, $\true$ and $\false$, are denoted \texttt{1} and \texttt{-1}.
Integer variables (including negative range values) are represented in
the order-encoding. When processing (bit-blasting) a declaration
$\mathtt{new\_int(I,Min,Max)}$, Prolog variable \texttt{I} is unified
with the tuple \texttt{(Min,Max,Bits,LastBit)} where \texttt{Min} and
\texttt{Max} are constants indicating the interval domain of
\texttt{I}, \texttt{Bits} is a list of $\mathtt{(Max-Min)}$ variables,
and \texttt{LastBit} is the last variable of \texttt{Bits}. This
representation is more concise than the one assumed for simplicity in
the previous sections and it also supports negative
numbers. Maintaining direct access to the last bit in the
representation (we already can access the first bit through the list
\texttt{Bits}) facilitates a (constant time) check if the lower and
upper bound values of a variable has changed. This way we can more
efficiently determine when (certain) simplification rules apply.
We make a few notes: (1) Integer variables must be declared before
use; (2) \bee\ allows the use of constants in constraints instead of
declaring them as integer variables (for example
$\mathtt{int\_gt(I,5)}$ represents a declaration
$\mathtt{new\_int(I',5,5)}$ together with the constraint
$\mathtt{int\_gt(I,I')}$); (3) integer variables can be negated.

\bee\ maintains constraints as a Prolog list
(of terms).  Each type of constraint is associated with corresponding
rules for simplification, decomposition, and encoding to CNF.
After bit-blasting, constraints are first simplified (equi-propagation
and partial evaluation) using these rules until no further rules
apply. During this process, if a pair of literals is equated (e.g.~ as
in \texttt{X=Y, X=-Y, X=1, X=-1}), then they are unified, thus
propagating the effect to other constraints. 
After constraint simplification, some constraints are decomposed, and
this process repeats.
We end up with a set of ``basic'' constraints (which cannot be further
decomposed or simplified).  These are then encoded to CNF.

\section{Experiments}\label{results}

We report on our experience in applying \bee. To appreciate the ease
in its use, and for further details, the reader is encouraged to view
the example encodings available with the tool \cite{bee2012}.
All experiments run on an Intel Core 2 Duo E8400 3.00GHz CPU with 4GB
memory under Linux (Ubuntu lucid, kernel 2.6.32-24-generic).
\bee\ is written in Prolog and run using SWI Prolog v6.0.2 64-bits.
Comparisons with Sugar (v1.15.0) are based on the use of identical
constraint models, apply the same SAT solver (CryptoMiniSat v2.5.1),
and run on the same machine.
For all of the tables describing
experiments, columns indicate:

\vspace{-3mm}
\qquad\begin{minipage}{0.45\linewidth}
\qquad\begin{itemize}
\item[\texttt{comp:}]     compile time (seconds) 
\item[\texttt{clauses:}]  number of CNF clauses
\end{itemize}
\end{minipage}
\begin{minipage}{0.4\linewidth}
\qquad\begin{itemize}
\item[\texttt{vars:}]     number of CNF variables
\item[\texttt{sat:}]      SAT solving time (seconds)
\end{itemize}
\end{minipage}
\medskip
%

We first focus on the impact of the dual representation for
\allDifferent\ constraints. We report on the application of \bee\ to
Quasi-group completion problems (QCP), proposed by
\citeN{DBLP:conf/cp/GomesSC97} as a constraint satisfaction benchmark,
where the model is a conjunction of \allDifferent\ constraints.

\vspace{-3mm}
\paragraph{\underline{Quasi-group completion:}}


\begin{table}[t]
\scriptsize
  \centering
    \begin{oldtabular}{lc|rrrr|rrrr|rrr}
\hline
    \multicolumn{2}{c|}{instance} & \multicolumn{4}{c|}{\bee\ (dual encoding)} & 
    \multicolumn{4}{c|}{\bee\ (order encoding)} & \multicolumn{3}{c}{Sugar} \\
\hline
               &       &comp &clauses  & vars  &  sat  &  comp &clauses&  vars &  sat   & clauses  &  vars &  sat  \\
\hline
     25-264-0  & sat   & 0.23  & 6509  & 1317  & 0.33  & 0.36  & 33224 & 887   & 8.95   & 126733 & 10770 & 34.20 \\
     25-264-1  & sat   & 0.20  & 7475  & 1508  & 3.29  & 0.30  & 34323 & 917   & 97.50  & 127222 & 10798 & 13.93 \\
     25-264-2  & sat   & 0.21  & 6531  & 1329  & 0.07  & 0.30  & 35238 & 905   & 2.46   & 127062 & 10787 & 8.06 \\
     25-264-3  & sat   & 0.21  & 6819  & 1374  & 0.83  & 0.29  & 32457 & 899   & 18.52  & 127757 & 10827 & 44.03 \\
     25-264-4  & sat   & 0.21  & 7082  & 1431  & 0.34  & 0.29  & 32825 & 897   & 19.08  & 126777 & 10779 & 85.92 \\
     25-264-5  & sat   & 0.21  & 7055  & 1431  & 3.12  & 0.30  & 33590 & 897   & 46.15  & 126973 & 10784 & 41.04 \\
     25-264-6  & sat   & 0.21  & 7712  & 1551  & 0.34  & 0.33  & 39015 & 932   & 69.81  & 128354 & 10850 & 12.67 \\
     25-264-7  & sat   & 0.21  & 7428  & 1496  & 0.13  & 0.30  & 36580 & 937   & 19.93  & 127106 & 10794 & 7.01 \\
     25-264-8  & sat   & 0.21  & 6603  & 1335  & 0.18  & 0.27  & 31561 & 896   & 10.32  & 124153 & 10687 & 9.69 \\
     25-264-9  & sat   & 0.21  & 6784  & 1350  & 0.19  & 0.27  & 35404 & 903   & 34.08  & 128423 & 10853 & 38.80 \\
    25-264-10  & unsat & 0.21  & 6491  & 1296  & 0.04  & 0.30  & 33321 & 930   & 10.92  & 126999 & 10785 & 57.75 \\
    25-264-11  & unsat & 0.12  & 1     & 0     & 0.00  & 0.28  & 37912 & 955   & 0.09   & 125373 & 10744 & 0.47 \\
    25-264-12  & unsat & 0.16  & 1     & 0     & 0.00  & 0.29  & 39135 & 984   & 0.08   & 127539 & 10815 & 0.57 \\
    25-264-13  & unsat & 0.12  & 1     & 0     & 0.00  & 0.29  & 35048 & 944   & 0.09   & 127026 & 10786 & 0.56 \\
    25-264-14  & unsat & 0.23  & 5984  & 1210  & 0.07  & 0.28  & 31093 & 885   & 11.60  & 126628 & 10771 & 15.93 \\
\hline
    Total      &       &       &       &       & 8.93  &       &       &       & 349.58 &       &       & 370.63 \\
\hline
    \end{oldtabular}
  \caption{QCP results for $25\times 25$ instances with 264 holes}
  \label{tab:qcpExpr}
\end{table}

We consider 15 instances from the 2008 CSP
competition\footnote{\url{http://www.cril.univ-artois.fr/CPAI08/}}.
Table~\ref{tab:qcpExpr} considers three settings: \bee\ with its dual
encoding for \allDifferent\ constraints, \bee\ using only the order
encoding (equivalent to using $\mathtt{int\_neq}$ constraints instead of
\allDifferent), and Sugar. The results indicate that:
(1) Application of \bee\ using the dual representation for
\allDifferent\ is 38 times faster and produces 20 times less clauses
(in average) than when using the order-encoding alone (despite the
need to maintain two encodings);
(2) Without the dual representation, solving encodings generated by
\bee\ is only slightly faster but \bee\ generates CNF encodings 4
times smaller (on average) than those generated by Sugar.
Observe that 3 instances are found unsatisfiable by \bee\ (indicated
by a CNF with a single clause and no variables). We comment that Sugar
preprocessing times are higher than those of \bee\ and not indicated
in the table.

\bigskip To further appreciate the impact of the tool we describe
results for three additional applications which shift the
state-of-the-art with respect to what could previously be solved.
%
The experiments clearly illustrate that \bee\ decreases the size of
CNF encodings as well as the subsequent SAT solving time.

\vspace{-3mm}
\paragraph{\underline{Magic labels:}}
In \cite{MacDougall2002} the authors conjecture that the $n$ vertex
complete graph, $K_n$, for $n\geq 5$ has a vertex magic total labeling
with magic constants for specific range of values of $k$, determined
by $n$. This conjecture is proved correct for all odd $n$ and verified
by brute force for $n=6$.
We address the cases for $n=8$ and $n=10$ which involve 15 instances
(different values of $K$) for $n=8$, and 23 (different values of $K$)
for $n=10$. Starting from the simple constraint model (illustrated by
the example in Figure~\ref{fig:vmtl-instance}), we add additional
constraints to exploit that the graphs are symmetric:
(1) We assume that the edge with the smallest label is $e_{1,2}$; 
(2) We assume that the labels of the edges incident to $v_1$ are
ordered and hence introduce constraints $e_{1,2} < e_{1,3} < \cdots <
e_{1,n}$; 
(3) We assume that the label of edge $e_{1,3}$ is smaller than the
labels of the edges incident to $v_2$ (except $e_{1,2}$) and introduce
constraints accordingly.
In this setting \bee\ can solve all except 2 instances with a 4 hour
timeout and Sugar can solve all except~4.

Table~\ref{tab:k8} depicts results for the 10 hardest instances for
$K_8$ and the 20 hardest for $K_{10}$ with a 4 hour time-out. \bee\
compilation times are on the order of 0.5 sec/instance for $K_8$ and
2.5 sec/instance for $K_{10}$. Sugar encoding times are slightly
larger. The instances are indicated by the magic constant, $k$; the
columns for \bee\ and Sugar indicate SAT solving times (in seconds).
The bottom two lines indicate average encoding sizes (numbers of
clauses and variables).

\begin{table}[t]
\scriptsize \hspace{-10mm}
  \begin{minipage}{0.37\linewidth}
    \begin{oldtabular}{c|c|r|r}
\hline
$K_8$& $k$   &  \multicolumn{1}{c}{\bee} &   \multicolumn{1}{c}{Sugar}  \\
\hline
&143 & 1.26 & 2.87\\
&142 & 10.14 & 1.62\\
&141 & 7.64 & 2.94\\
&140 & 14.68 & 6.46\\
&139 & 25.60 & 6.67\\
&138 & 12.99 & 2.80\\
&137 & 22.91 & 298.58\\
&136 & 14.46 & 251.82\\
&135 & 298.54 & 182.90\\
&134 & 331.80 & $\infty$\\
\hline 
\multicolumn{2}{c}{\vspace{-1mm}Average}\\
\hline
\multicolumn{2}{r|}{clauses {$\mathbf{\times 1000}$}} &248& 402\\
\multicolumn{2}{r|}{vars} &5688&9370\\
   \end{oldtabular} 
  \end{minipage}
  \begin{minipage}{0.26\linewidth}
    \begin{oldtabular}{c|c|r|r}
\hline
 $K_{10}$&k  & \multicolumn{1}{c}{\bee} &   \multicolumn{1}{c}{Sugar}  \\
\hline
&277 & 5.31 & 9.25\\
&276 & 7.11 & 9.91\\
&275 & 13.57 & 19.63\\
&274 & 4.93 & 9.24\\
&273 & 45.94 & 9.03\\
&272 & 22.74 & 86.45\\
&271 & 7.35 & 9.49\\
&270 & 6.03 & 55.94\\
&269 & 5.20 & 11.05\\
&268 & 94.44 & 424.89\\
\hline 
\multicolumn{2}{c}{\vspace{-1mm}~}\\
\hline
\multicolumn{4}{r}{clauses {$\mathbf{\times 1000}$}}\\
\multicolumn{4}{r}{vars}\\
\end{oldtabular}
\end{minipage}
  \begin{minipage}{0.25\linewidth}
\begin{oldtabular}{r|r|r}
\hline
  k~~  & \multicolumn{1}{c}{\bee} &   \multicolumn{1}{c}{Sugar}  \\
\hline
267~ & 88.51 & 175.70\\
266~ & 229.80 & 247.56\\
265~ & 1335.31 & 259.45\\
264~ & 486.09 & 513.61\\
263~ & 236.68 & 648.43\\
262~ & 1843.70 & 6429.25\\
261~ & 2771.60 & 7872.76\\
260~ & 4873.99 & $\infty$\\
259~ & $\infty$ & $\infty$\\
258~ & $\infty$ & $\infty$\\
\hline 
\multicolumn{2}{c}{\vspace{-1mm}Average}\\
\hline
\multicolumn{1}{r|}{}   &1229&1966 \\
\multicolumn{1}{r|}{}      &15529&25688 \\
\end{oldtabular}
\end{minipage}
  
\caption{VMTL results for $K_8$ and $K_{10}$ (times are in seconds)} 
  \label{tab:k8}
\end{table}

The results indicate that the Sugar encodings are (in average) about
60\% larger, while the average SAT solving time for the \bee\
encodings is about 2 times faster (average excluding instances
where Sugar times-out).

To address the two VMTL instances not solvable using the \bee\ models
described above ($K_{10}$ with magic labels 259 and 258), we partition
the problem fixing the values of $e_{1,2}$ and $e_{1,3}$ and
maintaining all of the other constraints. Analysis of the symmetry
breaking constraints indicates that this results in 198 new instances
for each of the two cases. The original VMTL instance is solved
if any one of of these 198 instances is solved. So, we solve them in
parallel. Fixing $e_{1,2}$ and $e_{1,3}$ ``fuels'' the compiler so the
encodings are considerably smaller.
The instance for $k= 259$ is solved in 1379.50 seconds where
$e_{1,2}=1$ and $e_{1,3}=6$. The compilation time is 2.09 seconds and
the encoding consists in 1056107 clauses and 14143 variables.

To the best of our knowledge, the hard instances from this suite are
beyond the reach of all previous approaches to program the search for
magic labels. The SAT based approach presented in \cite{Jaeger2010}
cannot handle these.\footnote{Personal communication (Gerold J\"ager),
  March 2012.} The comparison with Sugar indicates the impact of the
compiler.  


\vspace{-3mm}
\paragraph{\underline{DNA word problem:}}

\citeN{DBLP:journals/constraints/ManciniMPC08} provide a comparison of
several state-of-the-art solvers applied to the DNA word problem with
a variety of encoding techniques. Their best reported result is
a solution with 87 DNA words, obtained in 554 seconds, using an
OPL~\cite{opl} model with lexicographic order to break symmetry.
In~\cite{dnaWordPaper}, the authors report a solution composed from
two pairs of (t-part and m-part) sets $\tuple{T_1,M_1}$ and
$\tuple{T_2,M_2}$ where $|T_1|=6$, $|M_1|=16$, $|T_2|=2$,
$|M_2|=6$. This forms a set $S$ with $(6\times 16)+(2\times 6)=108$
DNA words.
Marc van Dongen reports a larger solution with 112
words.\footnote{See
  \url{http://www.cs.st-andrews.ac.uk/~ianm/CSPLib/}.}
Using \bee, we find, in a fraction of a second, a template of size 14
and a map of size 8. This provides a solution of size $14\times 8=112$
to the DNA word problem. Running Comet (v2.0.1) we find a 112 word
solution in about 10 seconds using a model by H\aa kan
Kjellerstrand.\footnote{See
  \url{http://www.hakank.org/comet/word_design_dna1.co}.}
We also prove that there does not exist a template of size 15 (0.15
seconds), nor a map of size 9 (4.47 seconds). These facts were unknown
prior to \bee.
Proving that there is no solution to the DNA word problem with
more than 112 words, not via the two part t-m strategy, is still an
open problem.

\vspace{-3mm}
\paragraph{\underline{Model Based Diagnostics}}

(MBD) is an artificial intelligence based
approach that aims to cope with the, so-called, diagnosis
problem~(e.g.~\cite{Reiter87}). In~\cite{MBD}, we (with other
researchers) focus on a notion of minimal cardinality MBD and apply
\bee\ to model and solve the instances of a standard MBD benchmark.
Experimental evidence (see \cite{MBD}), indicates that our approach is
superior to all existing algorithms for minimal cardinality MBD.  We
determine, for the first time, minimal cardinality diagnoses for the
entire standard benchmark.
Prior attempts to apply SAT for MBD (for example, by \citeN{Smith05}
and \citeN{Feldman10DX} where a MaxSAT solver is used) indicate that
SAT solvers perform poorly on the standard benchmarks.  So, \bee\
really makes the difference.

\section{Conclusion}

We introduce \bee, a compiler to encode finite domain constraints to
CNF. A key design point is to apply bit-level techniques, locally as
prescribed by the word-level constraints in a model.
Optimizations are based on equi-propagation and partial
evaluation.  Implemented in Prolog, compilation times are
typically small (measured in seconds) even for instances which result
in several millions of CNF clauses.  On the other hand, the reduction
in SAT solving time can be larger in orders of magnitude.

It is well-understood that making a CNF smaller is not the ultimate
goal: often smaller CNF's are harder to solve.  Indeed, one often
introduces redundancies to improve SAT encodings: so removing them is
counter productive.  Our experience is that \bee\ reduces the size of
an encoding in a way that is productive for the subsequent SAT
solving. In particular, by removing variables that can be determined
``at compile time'' to be definitely equal (or definitely different)
in any solution.

The simplification rules illustrated in Section~\ref{sec:compiling}
apply standard constraint programming techniques (i.e.~to reduce
variable domains). However, equi-propagation is more powerful. It
focuses, in general, in specializing the bit-level representation of
the constraints in view of equations implied by the constraints. In
this way it captures many of the well-known constraint programming
preprocessing techniques, and more.


Future work will investigate: how to strengthen the implementation of
equi-propagation using BDD's and SAT solving techniques, how to
improve the compiler implementation using better data-structures for
the constraint store (for example applying a CHR based approach for
the simplification rules), and how to enhance the underlying
constraint language.



\end{document}